\documentclass[conference]{IEEEtran}
\IEEEoverridecommandlockouts
\usepackage{cite}
\usepackage{amsmath,amssymb,amsfonts}
\usepackage{algorithmic}
\usepackage{graphicx}
\usepackage{textcomp}
\usepackage{xcolor}
\usepackage[unicode]{hyperref}
\usepackage{tikz}
\usepackage[center]{caption}
\usepackage{subcaption}
\def\BibTeX{{\rm B\kern-.05em{\sc i\kern-.025em b}\kern-.08em
    T\kern-.1667em\lower.7ex\hbox{E}\kern-.125emX}}
\begin{document}
	
\makeatletter
\newcommand{\linebreakand}{%
\end{@IEEEauthorhalign}
\hfill\mbox{}\par
\mbox{}\hfill\begin{@IEEEauthorhalign}
}
\makeatother

\title{RNNoise-Ex: Hybrid Speech Enhancement System\\
	 based on RNN and Spectral Features\\
}

\author{
\IEEEauthorblockN{Constantine C. Doumanidis}
\IEEEauthorblockA{
	\textit{School of Electrical and Computer Engineering} \\
\textit{Aristotle University of Thessaloniki}\\
Thessaloniki, Greece \\
kdoumani@ece.auth.gr}
\and
\IEEEauthorblockN{Christina Anagnostou}
\IEEEauthorblockA{
	\textit{School of Electrical and Computer Engineering} \\
	\textit{Aristotle University of Thessaloniki}\\
	Thessaloniki, Greece \\
cdanagnos@ece.auth.gr}
\linebreakand
\IEEEauthorblockN{Evangelia-Sofia Arvaniti}
\IEEEauthorblockA{
	\textit{School of Electrical and Computer Engineering} \\
	\textit{Aristotle University of Thessaloniki}\\
	Thessaloniki, Greece \\
earvaniti@ece.auth.gr}
\and
\IEEEauthorblockN{Anthi Papadopoulou}
\IEEEauthorblockA{
	\textit{School of Electrical and Computer Engineering} \\
	\textit{Aristotle University of Thessaloniki}\\
	Thessaloniki, Greece \\
anthipapado@ece.auth.gr}
{\footnotesize \textsuperscript{*}Note: Author name order was decided arbitrarily.}
}

\maketitle

\begin{abstract}
Recent interest in exploiting Deep Learning techniques for Noise Suppression, has led to the creation of Hybrid Denoising Systems that combine classic Signal Processing with Deep Learning. In this paper, we concentrated our efforts on extending the RNNoise denoising system with the inclusion of complementary features during the training phase. We present a comprehensive explanation of the set-up process of a modified system and present the comparative results derived from a performance evaluation analysis, using a reference version of RNNoise as control.
\end{abstract}


\section{Introduction}

Signal Processing has undoubtedly a wide range of useful applications in the modern world. Narrowing our focus on the domain of Audio Signal Processing, Speech Enhancement is an especially interesting subfield, due to the number of its applications, such as telecommunication networks, online video conferencing \cite{valin}, cochlear implants \cite{lai}, speech-to-text systems \cite{donahue}, etc. Speech Enhancement is heavily dependent on the concept of denoising; that is the removal of undesired audio signals that degrade the speech signal which may result in reduction of quality and intelligibility.

Noise Suppression is by no means a new field of study among scientists and engineers. The application, however, of modern techniques, ideas and innovations has enabled the field to grow and include some very promising denoising algorithms and systems. Such approaches can be divided to causal (e.g: \cite{valin}) and non-causal \cite{shifas}, depending on whether they exploit information in future signal frames to process the current. They can also be categorized in real time or non real time systems depending on their ability to process signal frames within a predefined time constraint. 

In the past, the focus of Noise Suppression was on the utilization of conventional signal processing techniques (filtering), which operate by estimating the statistical characteristics of the noise signal to be removed. Some commonly used such methods include Wiener \cite{weiner} and Kalman \cite{welchKalman} filters.

Following the increase of interest for machine learning shown in recent years by the scientific community, a new realm of possibility was now available to researchers in the vein of Noise Suppression. In the last decade especially, no small number of works have been published that approach the denoising problem by employing neural network architectures and innovative deep learning (DL) techniques to counteract non-stationary noise signals \cite{wang}. A yet more recent trend among researchers is the development of hybrid systems that combine both conventional and ML techniques. The motivations and advantages of such an approach appear to be:
\begin{itemize}
	\item the exploitation of existing knowledge on the problem nature, leading to the design of concept-aware systems
	\item engagement of data-driven approaches with large models that give the flexibility to better model the complex acoustic patterns of speech
	\item an increase of system performance by balancing/counteracting each method's weaknesses with the strengths of the other
	\item a decrease in unnecessary complexity as compared to purely ML techniques
	\item the better handling of auditory artifacts, which constitute one of the greatest hindrances in Speech Enhancement to date.
\end{itemize}

Elaborating on hybrid systems, in \cite{hybrid1} the noisy audio signal of the current time frame is first processed using
a suppression rule computed as a geometric mean of the clean speech estimation of the current frame using a conventional denoising technique and the result of the suppression rule
of the previous frame which was determined by an LSTM deep-learning technique. This first step is used to remove quasi-stationary noise components. The intermediate enhanced signal that results from the previously described process is then used to estimate the clean speech signal and the current frame suppression rule, using an LSTM-based approach. The aim of the second step is to efficiently remove non-stationary noise signals. The approach taken in \cite{hybrid2} follows a similar structure to that of \cite{hybrid1}. Namely, first the noisy signal is enhanced using the well-known Wiener filter. Afterwards, the resulting signal is further processed by a multi stream approach, which includes a number of denoising autoencoders and auto-associative memories, based on LSTM networks.

\subsection{Related Work: The RNNoise Implementation}

Another example of a system that combines both conventional and deep learning techniques, and the base of our work, is RNNoise \cite{valin}, implemented by Jean Marc Vallin with the support of Mozilla. RNNoise is a real-time system designed to run on simple hardware (e.g. Raspberry Pi). To achieve lower complexity, a Recurrent Neural Network (RNN) was employed for the portion of the spectral mask estimation process that was hard to tune and a conventional signal processing technique for the rest of it. In the following subsection we review some details of the RNNoise implementation that will better help the understanding of the work presented afterwards. 

In RNNoise, the denoising process is applied to 20 ms windows, overlapped by 50\% and windowed by a Vorbis window. For each window, follows the extraction of certain features that will be analyzed in Section II that are afterwards used as an input for the RNN. RNNoise operates on 48kHz full-band audio input. The network computes an ideal ratio mask (IRM) $m = [m_1, m_2, ..., m_{22}]^T \in \mathbb{R}^{22} : m_i \in [0, 1]$,  for 22 triangular bands derived from a modified version of the Bark scale that is similar to the Opus scale \cite{opus}. The 22 gains in $[\sqrt{m_1}, \sqrt{m_2}, ..., \sqrt{m_{22}}]^T$, after an interpolation, can be applied to the Discrete Fourier Transform (DFT) magnitudes of each window.

Before that, a pitch filter, namely a comb filter defined at the pitch period, is applied to each window. What this filter essentially implements is the addition of the original signal to its scaled and delayed by the pitch period version. The role of the pitch filter is to suppress noise between pitch harmonics of voiced speech, which is not feasible by the coarse 22-band gains mask produced by the neural network.

After the application of both the gains and the pitch filter, the waveform of the processed DFT is calculated and the overlap-add method is used to produce the final denoised signal. Practically, the overlap-add method is applied gradually, after each new 10 ms samples arrive, to achieve better response time. For a more detailed analysis of the RNNoise system see \cite{valin}.

\tikzset{every picture/.style={line width=0.75pt}} 
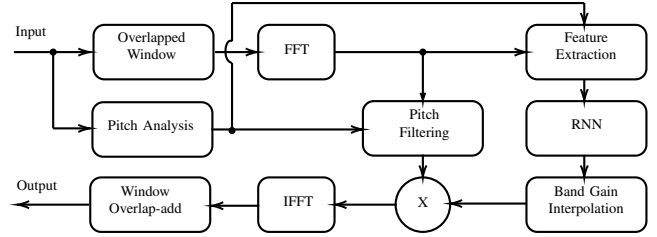
\begin{figure}[htbp]
	\begin{tikzpicture}[x=0.75pt,y=0.75pt,yscale=-1,xscale=1, thick,scale=0.55, every node/.style={scale=0.55}]
		
		\draw   (93,40) .. controls (93,34.48) and (97.48,30) .. (103,30) -- (193,30) .. controls (198.52,30) and (203,34.48) .. (203,40) -- (203,70) .. controls (203,75.52) and (198.52,80) .. (193,80) -- (103,80) .. controls (97.48,80) and (93,75.52) .. (93,70) -- cycle ;
		\draw   (92,111) .. controls (92,105.48) and (96.48,101) .. (102,101) -- (192,101) .. controls (197.52,101) and (202,105.48) .. (202,111) -- (202,141) .. controls (202,146.52) and (197.52,151) .. (192,151) -- (102,151) .. controls (96.48,151) and (92,146.52) .. (92,141) -- cycle ;
		\draw   (244,40) .. controls (244,34.48) and (248.48,30) .. (254,30) -- (304,30) .. controls (309.52,30) and (314,34.48) .. (314,40) -- (314,70) .. controls (314,75.52) and (309.52,80) .. (304,80) -- (254,80) .. controls (248.48,80) and (244,75.52) .. (244,70) -- cycle ;
		\draw   (490,40) .. controls (490,34.48) and (494.48,30) .. (500,30) -- (590,30) .. controls (595.52,30) and (600,34.48) .. (600,40) -- (600,70) .. controls (600,75.52) and (595.52,80) .. (590,80) -- (500,80) .. controls (494.48,80) and (490,75.52) .. (490,70) -- cycle ;
		\draw   (490,110) .. controls (490,104.48) and (494.48,100) .. (500,100) -- (590,100) .. controls (595.52,100) and (600,104.48) .. (600,110) -- (600,140) .. controls (600,145.52) and (595.52,150) .. (590,150) -- (500,150) .. controls (494.48,150) and (490,145.52) .. (490,140) -- cycle ;
		\draw   (340,110) .. controls (340,104.48) and (344.48,100) .. (350,100) -- (440,100) .. controls (445.52,100) and (450,104.48) .. (450,110) -- (450,140) .. controls (450,145.52) and (445.52,150) .. (440,150) -- (350,150) .. controls (344.48,150) and (340,145.52) .. (340,140) -- cycle ;
		\draw   (490,180) .. controls (490,174.48) and (494.48,170) .. (500,170) -- (590,170) .. controls (595.52,170) and (600,174.48) .. (600,180) -- (600,210) .. controls (600,215.52) and (595.52,220) .. (590,220) -- (500,220) .. controls (494.48,220) and (490,215.52) .. (490,210) -- cycle ;
		\draw   (370,195) .. controls (370,181.19) and (381.19,170) .. (395,170) .. controls (408.81,170) and (420,181.19) .. (420,195) .. controls (420,208.81) and (408.81,220) .. (395,220) .. controls (381.19,220) and (370,208.81) .. (370,195) -- cycle ;
		\draw   (244,180) .. controls (244,174.48) and (248.48,170) .. (254,170) -- (304,170) .. controls (309.52,170) and (314,174.48) .. (314,180) -- (314,210) .. controls (314,215.52) and (309.52,220) .. (304,220) -- (254,220) .. controls (248.48,220) and (244,215.52) .. (244,210) -- cycle ;
		\draw   (90,180) .. controls (90,174.48) and (94.48,170) .. (100,170) -- (190,170) .. controls (195.52,170) and (200,174.48) .. (200,180) -- (200,210) .. controls (200,215.52) and (195.52,220) .. (190,220) -- (100,220) .. controls (94.48,220) and (90,215.52) .. (90,210) -- cycle ;
		\draw    (20,55) -- (88,55) ;
		\draw [shift={(90,55)}, rotate = 180] [color={rgb, 255:red, 0; green, 0; blue, 0 }  ][line width=0.75]    (10.93,-3.29) .. controls (6.95,-1.4) and (3.31,-0.3) .. (0,0) .. controls (3.31,0.3) and (6.95,1.4) .. (10.93,3.29)   ;
		\draw    (201,126) -- (338,126) ;
		\draw [shift={(340,126)}, rotate = 180] [color={rgb, 255:red, 0; green, 0; blue, 0 }  ][line width=0.75]    (10.93,-3.29) .. controls (6.95,-1.4) and (3.31,-0.3) .. (0,0) .. controls (3.31,0.3) and (6.95,1.4) .. (10.93,3.29)   ;
		\draw    (314,55) -- (488,55) ;
		\draw [shift={(490,55)}, rotate = 180] [color={rgb, 255:red, 0; green, 0; blue, 0 }  ][line width=0.75]    (10.93,-3.29) .. controls (6.95,-1.4) and (3.31,-0.3) .. (0,0) .. controls (3.31,0.3) and (6.95,1.4) .. (10.93,3.29)   ;
		\draw    (395,55) -- (395,98) ;
		\draw [shift={(395,100)}, rotate = 270] [color={rgb, 255:red, 0; green, 0; blue, 0 }  ][line width=0.75]    (7.65,-2.3) .. controls (4.86,-0.97) and (2.31,-0.21) .. (0,0) .. controls (2.31,0.21) and (4.86,0.98) .. (7.65,2.3)   ;
		\draw [shift={(395,55)}, rotate = 90] [color={rgb, 255:red, 0; green, 0; blue, 0 }  ][fill={rgb, 255:red, 0; green, 0; blue, 0 }  ][line width=0.75]      (0, 0) circle [x radius= 2.34, y radius= 2.34]   ;
		\draw    (543,80) -- (543,98) ;
		\draw [shift={(543,100)}, rotate = 270] [color={rgb, 255:red, 0; green, 0; blue, 0 }  ][line width=0.75]    (10.93,-3.29) .. controls (6.95,-1.4) and (3.31,-0.3) .. (0,0) .. controls (3.31,0.3) and (6.95,1.4) .. (10.93,3.29)   ;
		\draw    (543,150) -- (543,168) ;
		\draw [shift={(543,170)}, rotate = 270] [color={rgb, 255:red, 0; green, 0; blue, 0 }  ][line width=0.75]    (10.93,-3.29) .. controls (6.95,-1.4) and (3.31,-0.3) .. (0,0) .. controls (3.31,0.3) and (6.95,1.4) .. (10.93,3.29)   ;
		\draw    (490,194) -- (422,194) ;
		\draw [shift={(420,194)}, rotate = 360] [color={rgb, 255:red, 0; green, 0; blue, 0 }  ][line width=0.75]    (10.93,-3.29) .. controls (6.95,-1.4) and (3.31,-0.3) .. (0,0) .. controls (3.31,0.3) and (6.95,1.4) .. (10.93,3.29)   ;
		\draw    (371,195) -- (316,195) ;
		\draw [shift={(314,195)}, rotate = 360] [color={rgb, 255:red, 0; green, 0; blue, 0 }  ][line width=0.75]    (10.93,-3.29) .. controls (6.95,-1.4) and (3.31,-0.3) .. (0,0) .. controls (3.31,0.3) and (6.95,1.4) .. (10.93,3.29)   ;
		\draw    (245,196) -- (203,196) ;
		\draw [shift={(201,196)}, rotate = 360] [color={rgb, 255:red, 0; green, 0; blue, 0 }  ][line width=0.75]    (10.93,-3.29) .. controls (6.95,-1.4) and (3.31,-0.3) .. (0,0) .. controls (3.31,0.3) and (6.95,1.4) .. (10.93,3.29)   ;
		\draw    (89,195) -- (25,195) ;
		\draw [shift={(23,195)}, rotate = 360] [color={rgb, 255:red, 0; green, 0; blue, 0 }  ][line width=0.75]    (10.93,-3.29) .. controls (6.95,-1.4) and (3.31,-0.3) .. (0,0) .. controls (3.31,0.3) and (6.95,1.4) .. (10.93,3.29)   ;
		\draw    (56,125) -- (88,125) ;
		\draw [shift={(90,125)}, rotate = 180] [color={rgb, 255:red, 0; green, 0; blue, 0 }  ][line width=0.75]    (10.93,-3.29) .. controls (6.95,-1.4) and (3.31,-0.3) .. (0,0) .. controls (3.31,0.3) and (6.95,1.4) .. (10.93,3.29)   ;
		\draw    (56,125) -- (56,55) ;
		\draw [shift={(56,55)}, rotate = 270] [color={rgb, 255:red, 0; green, 0; blue, 0 }  ][fill={rgb, 255:red, 0; green, 0; blue, 0 }  ][line width=0.75]      (0, 0) circle [x radius= 2.34, y radius= 2.34]   ;
		\draw    (204.5,55) -- (241.5,55) ;
		\draw [shift={(243.5,55)}, rotate = 180] [color={rgb, 255:red, 0; green, 0; blue, 0 }  ][line width=0.75]    (10.93,-3.29) .. controls (6.95,-1.4) and (3.31,-0.3) .. (0,0) .. controls (3.31,0.3) and (6.95,1.4) .. (10.93,3.29)   ;
		\draw    (395,149) -- (395,168) ;
		\draw [shift={(395,170)}, rotate = 270] [color={rgb, 255:red, 0; green, 0; blue, 0 }  ][line width=0.75]    (10.93,-3.29) .. controls (6.95,-1.4) and (3.31,-0.3) .. (0,0) .. controls (3.31,0.3) and (6.95,1.4) .. (10.93,3.29)   ;
		\draw    (220,65) -- (220,127) ;
		\draw [shift={(220,127)}, rotate = 90] [color={rgb, 255:red, 0; green, 0; blue, 0 }  ][fill={rgb, 255:red, 0; green, 0; blue, 0 }  ][line width=0.75]      (0, 0) circle [x radius= 2.34, y radius= 2.34]   ;
		\draw    (220,10) -- (220,45) ;
		\draw    (220,10) -- (543,10) ;
		\draw    (220,45) .. controls (213.67,41.33) and (210.33,62.67) .. (220,65) ;
		\draw    (543,10) -- (543,28) ;
		\draw [shift={(543,30)}, rotate = 270] [color={rgb, 255:red, 0; green, 0; blue, 0 }  ][line width=0.75]    (10.93,-3.29) .. controls (6.95,-1.4) and (3.31,-0.3) .. (0,0) .. controls (3.31,0.3) and (6.95,1.4) .. (10.93,3.29)   ;
		
		\draw (20,30) node [anchor=north west][inner sep=0.75pt]   [align=left] {Input};
		\draw (109,35) node [anchor=north west][inner sep=0.75pt]   [align=left] {\begin{minipage}[lt]{53.720000000000006pt}\setlength\topsep{0pt}
				\begin{center}
					Overlapped\\Window
				\end{center}
				
		\end{minipage}};
		\draw (101,116) node [anchor=north west][inner sep=0.75pt]   [align=left] {\begin{minipage}[lt]{64.60000000000001pt}\setlength\topsep{0pt}
				\begin{center}
					Pitch Analysis
				\end{center}
				
		\end{minipage}};
		\draw (264,46) node [anchor=north west][inner sep=0.75pt]   [align=left] {\begin{minipage}[lt]{21.080000000000002pt}\setlength\topsep{0pt}
				\begin{center}
					FFT
				\end{center}
				
		\end{minipage}};
		\draw (512,35) node [anchor=north west][inner sep=0.75pt]   [align=left] {\begin{minipage}[lt]{46.92pt}\setlength\topsep{0pt}
				\begin{center}
					Feature\\Extraction
				\end{center}
				
		\end{minipage}};
		\draw (528,115) node [anchor=north west][inner sep=0.75pt]   [align=left] {\begin{minipage}[lt]{23.8pt}\setlength\topsep{0pt}
				\begin{center}
					RNN
				\end{center}
				
		\end{minipage}};
		\draw (370,106) node [anchor=north west][inner sep=0.75pt]   [align=left] {\begin{minipage}[lt]{37.400000000000006pt}\setlength\topsep{0pt}
				\begin{center}
					Pitch\\Filtering
				\end{center}
				
		\end{minipage}};
		\draw (506,176) node [anchor=north west][inner sep=0.75pt]   [align=left] {\begin{minipage}[lt]{55.760000000000005pt}\setlength\topsep{0pt}
				\begin{center}
					Band Gain\\Interpolation
				\end{center}
				
		\end{minipage}};
		\draw (389,187) node [anchor=north west][inner sep=0.75pt]   [align=left] {X};
		\draw (264,186) node [anchor=north west][inner sep=0.75pt]   [align=left] {\begin{minipage}[lt]{23.12pt}\setlength\topsep{0pt}
				\begin{center}
					IFFT
				\end{center}
				
		\end{minipage}};
		\draw (101,176) node [anchor=north west][inner sep=0.75pt]   [align=left] {\begin{minipage}[lt]{57.120000000000005pt}\setlength\topsep{0pt}
				\begin{center}
					Window\\Overlap-add
				\end{center}
				
		\end{minipage}};
		\draw (21,171) node [anchor=north west][inner sep=0.75pt]   [align=left] {Output};

	\end{tikzpicture}
	
	\caption{RNNoise system architecture overview \cite{valin}.}
	\label{figRNNoiseArchitecture}
	
\end{figure}

Now that the basics of RNNoise and similar systems have been covered, what ensues is the presentation of our modifications to the system. In Section II, the input features –new and old- are described, as well as the training and evaluation datasets and toolchains. The assessment of the results of the new system, as well as a comparison with a retrained, reference version of the original RNNoise system, along with some comments, comprise Section III of the paper. Finally, in Section IV we conclude and summarize everything discussed in the previous sections.

\section{Methodology}

The main objective in this study is to explore possible performance gains over the original RNNoise system \cite{valin} by modifying it so that it utilizes extended information regarding its input. We first review the input features, then train a reference RNNoise system and our extended system using our selected datasets, so that we can later evaluate them and make a fair comparison between the two, and finally present the toolchain we developed to aid us in this process.

The original system by Valin (2018) uses 42 input features to perform speech enhancement \cite{valin}. The first 22 are Bark Frequency Cepstral Coefficients (BFCCs) as derived from applying the Discreet Cosine Transformation (DCT) on the log spectrum of the previously mentioned modified Bark scale. The next 12 features are the first and second order temporal derivatives of the first 6 BFCCs. The following 6 features are calculated by applying the DCT on the pitch correlation across frequency bands and selecting the first 6 coefficients. The final two features are the pitch period and a spectral non-stationary metric that assists in speech detection.

Given that the original system generally relies upon features related to pitch and BFCCs, we decided to explore the potential of combining them with characteristics of a different nature. Reviewing commonly used features in the literature \cite{giannakopoulos} \cite{andersson} \cite{maningo} \cite{peeters} \cite{dubnov}, we chose to use the following, standardized to zero mean and unit variance, for the full spectral and temporal range of each 20 ms frame processed by the extended system: 

\begin{itemize}
	\item \textbf{Spectral Centroid:} Signal's spectral ``center of mass''
	\item \textbf{Spectral Bandwidth:} Signal's highest minus lowest frequency
	\item \textbf{Spectral Roll-Off:} Threshold frequency over which 90\% of the signal's energy is situated
\end{itemize}

To calculate Spectral Centroid, first the Discrete Fourier Transform (DFT) for each frame is calculated using \eqref{eqAnk}, where $k$ is the $k$-th frequency for the $n$-th frame, $x(m)$ is the input signal, $w(m)$ is a window function and $L$ is the window's length. Spectral Centroid is then calculated using \eqref{eqSpectralCentroid} with $K$ being the DFT's order.
\begin{equation}\label{eqAnk}
	A(n,k) = |\sum_{m=-\infty}^{\infty} x(m)w(nL-M)e^{-j(\frac{2\pi}{L}km)}|
\end{equation}

\begin{equation}\label{eqSpectralCentroid}
	SC(n) = \frac{\sum_{k=0}^{K-1}k\cdot|A(n,k)|^2}{\sum_{k=0}^{K-1}|A(n,k)|^2}
\end{equation}

Spectral Roll-Off is calculated using \eqref{eqRollOff} where $N$ is the total number of frames, $K$ is the order of the DFT, $TH$  is a threshold (usually $\approx0.9$) and $A(n,k)$ is calculated using \eqref{eqAnk}.
\begin{equation}\label{eqRollOff}
	SRF(n) = max(h|\sum_{k=0}^{h}A(n,k) < TH\cdot\sum_{k=0}^{K-1}|A(n,k)|^2)
\end{equation}

To train and evaluate our extended system we implemented a modified toolchain which reuses modified parts of \cite{valin}. In the following paragraphs we present the components of our toolchain for feature extraction, training and evaluation. The full training and evaluation toolchain is visualized in Fig.~\ref{figToolchain}. Our source code is publicly available \footnote{Source Code: \url{https://github.com/CedArctic/rnnoise-ex}}.

\tikzset{every picture/.style={line width=0.75pt}} 
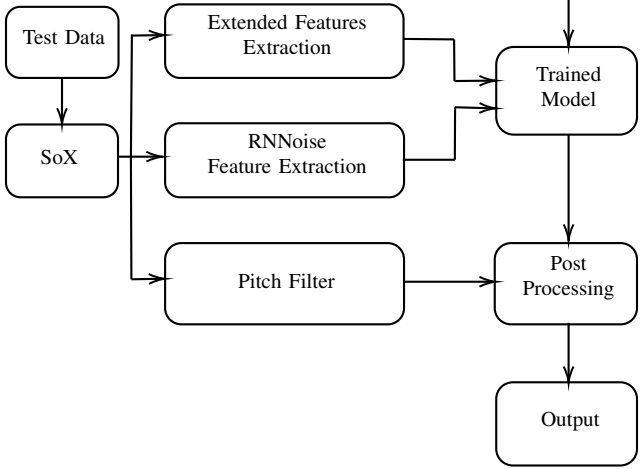
\begin{figure}[htbp]
	\begin{tikzpicture}[x=0.75pt,y=0.75pt,yscale=-1,xscale=1, thick,scale=0.75, every node/.style={scale=0.8}]
		
		\draw   (22,261.9) .. controls (22,256.62) and (26.28,252.33) .. (31.57,252.33) -- (88.1,252.33) .. controls (93.38,252.33) and (97.67,256.62) .. (97.67,261.9) -- (97.67,290.6) .. controls (97.67,295.88) and (93.38,300.17) .. (88.1,300.17) -- (31.57,300.17) .. controls (26.28,300.17) and (22,295.88) .. (22,290.6) -- cycle ;
		\draw   (22,341.3) .. controls (22,335.89) and (26.39,331.5) .. (31.8,331.5) -- (87.53,331.5) .. controls (92.95,331.5) and (97.33,335.89) .. (97.33,341.3) -- (97.33,370.7) .. controls (97.33,376.11) and (92.95,380.5) .. (87.53,380.5) -- (31.8,380.5) .. controls (26.39,380.5) and (22,376.11) .. (22,370.7) -- cycle ;
		\draw    (106.33,435.67) -- (106,353.33) ;
		\draw    (106.33,435.67) -- (127,435.36) ;
		\draw [shift={(129,435.33)}, rotate = 539.1600000000001] [color={rgb, 255:red, 0; green, 0; blue, 0 }  ][line width=0.75]    (10.93,-3.29) .. controls (6.95,-1.4) and (3.31,-0.3) .. (0,0) .. controls (3.31,0.3) and (6.95,1.4) .. (10.93,3.29)   ;
		\draw    (97.33,353.17) -- (125.33,353.17) ;
		\draw [shift={(127.33,353.17)}, rotate = 180] [color={rgb, 255:red, 0; green, 0; blue, 0 }  ][line width=0.75]    (10.93,-3.29) .. controls (6.95,-1.4) and (3.31,-0.3) .. (0,0) .. controls (3.31,0.3) and (6.95,1.4) .. (10.93,3.29)   ;
		\draw   (129,260.57) .. controls (129,255.1) and (133.43,250.67) .. (138.9,250.67) -- (279.43,250.67) .. controls (284.9,250.67) and (289.33,255.1) .. (289.33,260.57) -- (289.33,290.27) .. controls (289.33,295.73) and (284.9,300.17) .. (279.43,300.17) -- (138.9,300.17) .. controls (133.43,300.17) and (129,295.73) .. (129,290.27) -- cycle ;
		\draw    (59.67,300.17) -- (59.67,329.5) ;
		\draw [shift={(59.67,331.5)}, rotate = 270] [color={rgb, 255:red, 0; green, 0; blue, 0 }  ][line width=0.75]    (10.93,-3.29) .. controls (6.95,-1.4) and (3.31,-0.3) .. (0,0) .. controls (3.31,0.3) and (6.95,1.4) .. (10.93,3.29)   ;
		\draw   (351.67,516.2) .. controls (351.67,510.01) and (356.68,505) .. (362.87,505) -- (435.13,505) .. controls (441.32,505) and (446.33,510.01) .. (446.33,516.2) -- (446.33,549.8) .. controls (446.33,555.99) and (441.32,561) .. (435.13,561) -- (362.87,561) .. controls (356.68,561) and (351.67,555.99) .. (351.67,549.8) -- cycle ;
		\draw   (350.67,422.27) .. controls (350.67,416.23) and (355.56,411.33) .. (361.6,411.33) -- (435.4,411.33) .. controls (441.44,411.33) and (446.33,416.23) .. (446.33,422.27) -- (446.33,455.07) .. controls (446.33,461.1) and (441.44,466) .. (435.4,466) -- (361.6,466) .. controls (355.56,466) and (350.67,461.1) .. (350.67,455.07) -- cycle ;
		\draw    (400.33,465) -- (400.33,503.33) ;
		\draw [shift={(400.33,505.33)}, rotate = 270] [color={rgb, 255:red, 0; green, 0; blue, 0 }  ][line width=0.75]    (10.93,-3.29) .. controls (6.95,-1.4) and (3.31,-0.3) .. (0,0) .. controls (3.31,0.3) and (6.95,1.4) .. (10.93,3.29)   ;
		\draw  [dash pattern={on 0.84pt off 2.51pt}]  (22,190) -- (446.33,190) ;
		\draw   (129,340.63) .. controls (129,335.13) and (133.46,330.67) .. (138.97,330.67) -- (279.7,330.67) .. controls (285.2,330.67) and (289.67,335.13) .. (289.67,340.63) -- (289.67,370.53) .. controls (289.67,376.04) and (285.2,380.5) .. (279.7,380.5) -- (138.97,380.5) .. controls (133.46,380.5) and (129,376.04) .. (129,370.53) -- cycle ;
		\draw   (129,422.27) .. controls (129,416.23) and (133.9,411.33) .. (139.93,411.33) -- (278.73,411.33) .. controls (284.77,411.33) and (289.67,416.23) .. (289.67,422.27) -- (289.67,455.07) .. controls (289.67,461.1) and (284.77,466) .. (278.73,466) -- (139.93,466) .. controls (133.9,466) and (129,461.1) .. (129,455.07) -- cycle ;
		\draw    (288.83,355.25) -- (323.83,355.25) ;
		\draw    (323.58,302.25) -- (323.33,274.25) ;
		\draw    (289.33,273.92) -- (323.33,274.25) ;
		\draw   (351.67,291.13) .. controls (351.67,284.62) and (356.95,279.33) .. (363.47,279.33) -- (434.53,279.33) .. controls (441.05,279.33) and (446.33,284.62) .. (446.33,291.13) -- (446.33,326.53) .. controls (446.33,333.05) and (441.05,338.33) .. (434.53,338.33) -- (363.47,338.33) .. controls (356.95,338.33) and (351.67,333.05) .. (351.67,326.53) -- cycle ;
		\draw    (323.58,302.25) -- (349.33,302.02) ;
		\draw [shift={(351.33,302)}, rotate = 539.48] [color={rgb, 255:red, 0; green, 0; blue, 0 }  ][line width=0.75]    (10.93,-3.29) .. controls (6.95,-1.4) and (3.31,-0.3) .. (0,0) .. controls (3.31,0.3) and (6.95,1.4) .. (10.93,3.29)   ;
		\draw    (400.33,338) -- (400.33,407.67) ;
		\draw [shift={(400.33,409.67)}, rotate = 270] [color={rgb, 255:red, 0; green, 0; blue, 0 }  ][line width=0.75]    (10.93,-3.29) .. controls (6.95,-1.4) and (3.31,-0.3) .. (0,0) .. controls (3.31,0.3) and (6.95,1.4) .. (10.93,3.29)   ;
		\draw    (323.83,355.25) -- (324,319.92) ;
		\draw    (324,319.92) -- (349.33,320.61) ;
		\draw [shift={(351.33,320.67)}, rotate = 181.57] [color={rgb, 255:red, 0; green, 0; blue, 0 }  ][line width=0.75]    (10.93,-3.29) .. controls (6.95,-1.4) and (3.31,-0.3) .. (0,0) .. controls (3.31,0.3) and (6.95,1.4) .. (10.93,3.29)   ;
		\draw    (289.67,437.33) -- (349.33,437.33) ;
		\draw [shift={(351.33,437.33)}, rotate = 180] [color={rgb, 255:red, 0; green, 0; blue, 0 }  ][line width=0.75]    (10.93,-3.29) .. controls (6.95,-1.4) and (3.31,-0.3) .. (0,0) .. controls (3.31,0.3) and (6.95,1.4) .. (10.93,3.29)   ;
		\draw   (22,51.9) .. controls (22,46.62) and (26.28,42.33) .. (31.57,42.33) -- (88.1,42.33) .. controls (93.38,42.33) and (97.67,46.62) .. (97.67,51.9) -- (97.67,80.6) .. controls (97.67,85.88) and (93.38,90.17) .. (88.1,90.17) -- (31.57,90.17) .. controls (26.28,90.17) and (22,85.88) .. (22,80.6) -- cycle ;
		\draw   (22,131.3) .. controls (22,125.89) and (26.39,121.5) .. (31.8,121.5) -- (87.53,121.5) .. controls (92.95,121.5) and (97.33,125.89) .. (97.33,131.3) -- (97.33,160.7) .. controls (97.33,166.11) and (92.95,170.5) .. (87.53,170.5) -- (31.8,170.5) .. controls (26.39,170.5) and (22,166.11) .. (22,160.7) -- cycle ;
		\draw   (129,50.57) .. controls (129,45.1) and (133.43,40.67) .. (138.9,40.67) -- (279.43,40.67) .. controls (284.9,40.67) and (289.33,45.1) .. (289.33,50.57) -- (289.33,80.27) .. controls (289.33,85.73) and (284.9,90.17) .. (279.43,90.17) -- (138.9,90.17) .. controls (133.43,90.17) and (129,85.73) .. (129,80.27) -- cycle ;
		\draw    (58.67,90.17) -- (58.67,119.5) ;
		\draw [shift={(58.67,121.5)}, rotate = 270] [color={rgb, 255:red, 0; green, 0; blue, 0 }  ][line width=0.75]    (10.93,-3.29) .. controls (6.95,-1.4) and (3.31,-0.3) .. (0,0) .. controls (3.31,0.3) and (6.95,1.4) .. (10.93,3.29)   ;
		\draw    (210.33,121.33) -- (209.71,93) ;
		\draw [shift={(209.67,91)}, rotate = 448.74] [color={rgb, 255:red, 0; green, 0; blue, 0 }  ][line width=0.75]    (10.93,-3.29) .. controls (6.95,-1.4) and (3.31,-0.3) .. (0,0) .. controls (3.31,0.3) and (6.95,1.4) .. (10.93,3.29)   ;
		\draw   (129,130.63) .. controls (129,125.13) and (133.46,120.67) .. (138.97,120.67) -- (279.7,120.67) .. controls (285.2,120.67) and (289.67,125.13) .. (289.67,130.63) -- (289.67,160.53) .. controls (289.67,166.04) and (285.2,170.5) .. (279.7,170.5) -- (138.97,170.5) .. controls (133.46,170.5) and (129,166.04) .. (129,160.53) -- cycle ;
		\draw    (323.58,92.25) -- (323.33,64.25) ;
		\draw   (351.67,81.13) .. controls (351.67,74.62) and (356.95,69.33) .. (363.47,69.33) -- (434.53,69.33) .. controls (441.05,69.33) and (446.33,74.62) .. (446.33,81.13) -- (446.33,116.53) .. controls (446.33,123.05) and (441.05,128.33) .. (434.53,128.33) -- (363.47,128.33) .. controls (356.95,128.33) and (351.67,123.05) .. (351.67,116.53) -- cycle ;
		\draw    (323.58,92.25) -- (349.33,92.02) ;
		\draw [shift={(351.33,92)}, rotate = 539.48] [color={rgb, 255:red, 0; green, 0; blue, 0 }  ][line width=0.75]    (10.93,-3.29) .. controls (6.95,-1.4) and (3.31,-0.3) .. (0,0) .. controls (3.31,0.3) and (6.95,1.4) .. (10.93,3.29)   ;
		\draw    (400.33,128) -- (400.33,276.67) ;
		\draw [shift={(400.33,278.67)}, rotate = 270] [color={rgb, 255:red, 0; green, 0; blue, 0 }  ][line width=0.75]    (10.93,-3.29) .. controls (6.95,-1.4) and (3.31,-0.3) .. (0,0) .. controls (3.31,0.3) and (6.95,1.4) .. (10.93,3.29)   ;
		\draw    (323.83,145.25) -- (324,109.92) ;
		\draw    (324,109.92) -- (349.33,110.61) ;
		\draw [shift={(351.33,110.67)}, rotate = 181.57] [color={rgb, 255:red, 0; green, 0; blue, 0 }  ][line width=0.75]    (10.93,-3.29) .. controls (6.95,-1.4) and (3.31,-0.3) .. (0,0) .. controls (3.31,0.3) and (6.95,1.4) .. (10.93,3.29)   ;
		\draw    (289.33,63.92) -- (323.33,64.25) ;
		\draw    (289.67,144.58) -- (323.67,144.92) ;
		\draw    (98,144.5) -- (126,144.5) ;
		\draw [shift={(128,144.5)}, rotate = 180] [color={rgb, 255:red, 0; green, 0; blue, 0 }  ][line width=0.75]    (10.93,-3.29) .. controls (6.95,-1.4) and (3.31,-0.3) .. (0,0) .. controls (3.31,0.3) and (6.95,1.4) .. (10.93,3.29)   ;
		\draw    (106,353.33) -- (106.67,271) ;
		\draw    (106.33,271.67) -- (127,271.36) ;
		\draw [shift={(129,271.33)}, rotate = 539.1600000000001] [color={rgb, 255:red, 0; green, 0; blue, 0 }  ][line width=0.75]    (10.93,-3.29) .. controls (6.95,-1.4) and (3.31,-0.3) .. (0,0) .. controls (3.31,0.3) and (6.95,1.4) .. (10.93,3.29)   ;
		
		\draw (27.33,266.47) node [anchor=north west][inner sep=0.75pt]   [align=left] {\begin{minipage}[lt]{44.88pt}\setlength\topsep{0pt}
				\begin{center}
					Test Data
				\end{center}
				
		\end{minipage}};
		\draw (43.67,346.67) node [anchor=north west][inner sep=0.75pt]   [align=left] {SoX};
		\draw (148.1,255.67) node [anchor=north west][inner sep=0.75pt]   [align=left] {\begin{minipage}[lt]{86.36pt}\setlength\topsep{0pt}
				\begin{center}
					Extended Features\\Extraction
				\end{center}
				
		\end{minipage}};
		\draw (173.67,430.33) node [anchor=north west][inner sep=0.75pt]   [align=left] {\begin{minipage}[lt]{49.64pt}\setlength\topsep{0pt}
				\begin{center}
					Pitch Filter
				\end{center}
				
		\end{minipage}};
		\draw (376.69,523.33) node [anchor=north west][inner sep=0.75pt]   [align=left] {\begin{minipage}[lt]{32.64pt}\setlength\topsep{0pt}
				\begin{center}
					Output
				\end{center}
				
		\end{minipage}};
		\draw (361,418.33) node [anchor=north west][inner sep=0.75pt]   [align=left] {\begin{minipage}[lt]{53.040000000000006pt}\setlength\topsep{0pt}
				\begin{center}
					Post\\Processing
				\end{center}
				
		\end{minipage}};
		\draw (23,4.67) node [anchor=north west][inner sep=0.75pt]  [font=\Large] [align=left] {\textbf{Training}};
	\draw (25,213.67) node [anchor=north west][inner sep=0.75pt]  [font=\Large] [align=left] {\textbf{Evaluation}};
	\draw (150.1,336) node [anchor=north west][inner sep=0.75pt]   [align=left] {\begin{minipage}[lt]{83.64pt}\setlength\topsep{0pt}
		\begin{center}
			RNNoise\\Feature Extraction
		\end{center}
		
	\end{minipage}};
	\draw (373.67,290.67) node [anchor=north west][inner sep=0.75pt]   [align=left] {\begin{minipage}[lt]{35.36pt}\setlength\topsep{0pt}
		\begin{center}
			Trained\\Model
		\end{center}
		
	\end{minipage}};
	\draw (33.33,47.47) node [anchor=north west][inner sep=0.75pt]   [align=left] {\begin{minipage}[lt]{37.400000000000006pt}\setlength\topsep{0pt}
		\begin{center}
			Training\\Data
		\end{center}
		
	\end{minipage}};
	\draw (148.1,45.67) node [anchor=north west][inner sep=0.75pt]   [align=left] {\begin{minipage}[lt]{86.36pt}\setlength\topsep{0pt}
		\begin{center}
			Extended Features\\Extraction
		\end{center}
		
	\end{minipage}};
	\draw (150.1,126) node [anchor=north west][inner sep=0.75pt]   [align=left] {\begin{minipage}[lt]{83.64pt}\setlength\topsep{0pt}
		\begin{center}
			RNNoise Mixing \&\\Feature Extraction
		\end{center}
		
	\end{minipage}};
	\draw (372,91.17) node [anchor=north west][inner sep=0.75pt]   [align=left] {\begin{minipage}[lt]{37.400000000000006pt}\setlength\topsep{0pt}
		\begin{center}
			Training
		\end{center}
		
	\end{minipage}};
	\draw (44.33,134.67) node [anchor=north west][inner sep=0.75pt]   [align=left] {SoX};

	\end{tikzpicture}

	\caption{Training and Evaluation toolchain overview.}
	\label{figToolchain}
\end{figure}

For feature extraction we first use Sound eXchange (SoX) \cite{sox} to concatenate and convert the input clean speech and noise to RAW format files which we then process using the appropriate tool from \cite{valin} to generate the training samples, by mixing clean speech and noise tracks as shown in \cite{valin}, and extract the original 42 features as well as additional features used for training. After, we process the training samples using a feature extraction tool to extract the additional features.

We train the extended system using Keras with Tensorflow \cite{tensorflow} through the training tool which we modified according to the extended system's parameters. Both reference and extended system are trained through the course of 120 epochs with 8 steps each using the Adam optimizer with the learning rate set to 0.001. We use the loss function \eqref{eqLoss} (as proposed in \cite{valin}), where $m$ is the ground truth IRM mask, $\hat{m}$ is the mask calculated by the RNN, $\gamma = \frac{1}{2}$ is a parameter that tunes the suppression's aggressiveness and $N$ is the number of bands, which in our case is l to 22. During training, both systems process 3 600 000 audio frames, each with a non-overlapping 10 ms duration.

	\begin{multline}\label{eqLoss}
		L(m, \hat{m}) =\\
			\shoveright{\frac{1}{N} \cdot \bigg(
			10 \cdot \sum_{i=1}^{N}\Big(
				min(m_i + 1, 1) \cdot (10\cdot(m_i-\hat{m_i})^4 \\
				+ (\sqrt{\hat{m_i}} - \sqrt{m_i} )^\gamma 
				- 0.01 \cdot m_i \cdot log(\hat{m_i}))
			\Big)}\\
			- \frac{1}{2} \cdot \sum_{i=1}^{N}\Big(
				2 \cdot |m_i-0.5| \cdot m_i \cdot log(\hat{m_i})
			\Big)
		\bigg)
	\end{multline}

To evaluate inputs to our trained extended system, we first extract the 42 features presented in \cite{valin} using the original feature extraction tool and then merge them with the additional features extracted using the tools previously described. We pass this data along with the input audio file to the evaluation tool which calculates, interpolates and applies the modified Bark scale gains along with a pitch filter to the audio file as described in \cite{valin}.

The architecture of the neural network used follows that of the original RNNoise system with the difference that the input layer creates a tensor whose size is modified to fit that of the increased number of features. The topology is presented in detail in Fig.~\ref{figRNNtopology} and the system contains 215 units and 4 hidden layers.

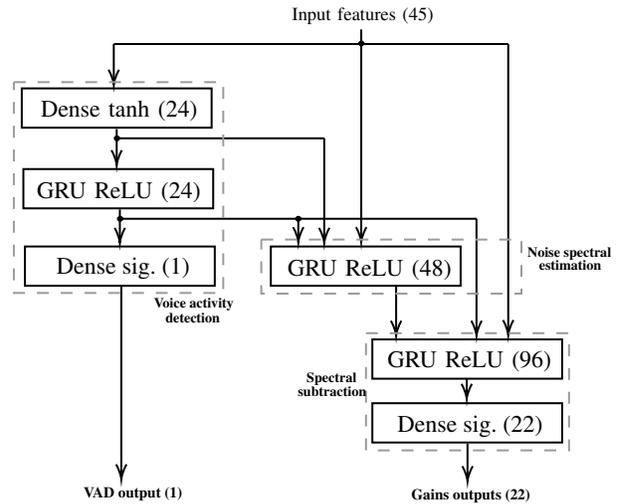
\begin{figure}[htbp]
	
	\tikzset{every picture/.style={line width=0.75pt}} 
	\begin{tikzpicture}[x=0.75pt,y=0.75pt,yscale=-0.7,xscale=0.8]
		
		\draw   (50,68) -- (169.67,68) -- (169.67,96.33) -- (50,96.33) -- cycle ;
		\draw   (51,126.17) -- (170.67,126.17) -- (170.67,154.5) -- (51,154.5) -- cycle ;
		\draw   (52,182) -- (171.67,182) -- (171.67,210.33) -- (52,210.33) -- cycle ;
		\draw   (207,182) -- (326.67,182) -- (326.67,210.33) -- (207,210.33) -- cycle ;
		\draw   (271,249) -- (390.67,249) -- (390.67,277.33) -- (271,277.33) -- cycle ;
		\draw   (271,295) -- (390.67,295) -- (390.67,323.33) -- (271,323.33) -- cycle ;
		\draw    (108,36) -- (356.67,36) ;
		\draw    (108,36) -- (108,63.5) ;
		\draw [shift={(108,65.5)}, rotate = 270] [color={rgb, 255:red, 0; green, 0; blue, 0 }  ][line width=0.75]    (10.93,-3.29) .. controls (6.95,-1.4) and (3.31,-0.3) .. (0,0) .. controls (3.31,0.3) and (6.95,1.4) .. (10.93,3.29)   ;
		\draw    (356.67,36) -- (356.67,244) ;
		\draw [shift={(356.67,246)}, rotate = 270] [color={rgb, 255:red, 0; green, 0; blue, 0 }  ][line width=0.75]    (10.93,-3.29) .. controls (6.95,-1.4) and (3.31,-0.3) .. (0,0) .. controls (3.31,0.3) and (6.95,1.4) .. (10.93,3.29)   ;
		\draw    (109.83,97.33) -- (109.83,121.5) ;
		\draw [shift={(109.83,123.5)}, rotate = 270] [color={rgb, 255:red, 0; green, 0; blue, 0 }  ][line width=0.75]    (10.93,-3.29) .. controls (6.95,-1.4) and (3.31,-0.3) .. (0,0) .. controls (3.31,0.3) and (6.95,1.4) .. (10.93,3.29)   ;
		\draw    (110,104) -- (240.67,104) ;
		\draw    (111.83,155.33) -- (111.83,177.5) ;
		\draw [shift={(111.83,179.5)}, rotate = 270] [color={rgb, 255:red, 0; green, 0; blue, 0 }  ][line width=0.75]    (10.93,-3.29) .. controls (6.95,-1.4) and (3.31,-0.3) .. (0,0) .. controls (3.31,0.3) and (6.95,1.4) .. (10.93,3.29)   ;
		\draw    (240.67,104) -- (240.67,177.5) ;
		\draw [shift={(240.67,179.5)}, rotate = 270] [color={rgb, 255:red, 0; green, 0; blue, 0 }  ][line width=0.75]    (10.93,-3.29) .. controls (6.95,-1.4) and (3.31,-0.3) .. (0,0) .. controls (3.31,0.3) and (6.95,1.4) .. (10.93,3.29)   ;
		\draw    (112,162) -- (336.67,162) ;
		\draw    (336.67,162) -- (336.67,244.5) ;
		\draw [shift={(336.67,246.5)}, rotate = 270] [color={rgb, 255:red, 0; green, 0; blue, 0 }  ][line width=0.75]    (10.93,-3.29) .. controls (6.95,-1.4) and (3.31,-0.3) .. (0,0) .. controls (3.31,0.3) and (6.95,1.4) .. (10.93,3.29)   ;
		\draw    (224.33,162) -- (224.33,177) ;
		\draw [shift={(224.33,179)}, rotate = 270] [color={rgb, 255:red, 0; green, 0; blue, 0 }  ][line width=0.75]    (10.93,-3.29) .. controls (6.95,-1.4) and (3.31,-0.3) .. (0,0) .. controls (3.31,0.3) and (6.95,1.4) .. (10.93,3.29)   ;
		\draw    (264,25.5) -- (264,177) ;
		\draw [shift={(264,179)}, rotate = 270] [color={rgb, 255:red, 0; green, 0; blue, 0 }  ][line width=0.75]    (10.93,-3.29) .. controls (6.95,-1.4) and (3.31,-0.3) .. (0,0) .. controls (3.31,0.3) and (6.95,1.4) .. (10.93,3.29)   ;
		\draw    (286,210) -- (286,244.5) ;
		\draw [shift={(286,246.5)}, rotate = 270] [color={rgb, 255:red, 0; green, 0; blue, 0 }  ][line width=0.75]    (10.93,-3.29) .. controls (6.95,-1.4) and (3.31,-0.3) .. (0,0) .. controls (3.31,0.3) and (6.95,1.4) .. (10.93,3.29)   ;
		\draw    (330.5,277) -- (330.5,290) ;
		\draw [shift={(330.5,292)}, rotate = 270] [color={rgb, 255:red, 0; green, 0; blue, 0 }  ][line width=0.75]    (10.93,-3.29) .. controls (6.95,-1.4) and (3.31,-0.3) .. (0,0) .. controls (3.31,0.3) and (6.95,1.4) .. (10.93,3.29)   ;
		\draw    (331,324) -- (331,349.33) ;
		\draw [shift={(331,351.33)}, rotate = 270] [color={rgb, 255:red, 0; green, 0; blue, 0 }  ][line width=0.75]    (10.93,-3.29) .. controls (6.95,-1.4) and (3.31,-0.3) .. (0,0) .. controls (3.31,0.3) and (6.95,1.4) .. (10.93,3.29)   ;
		\draw  [fill={rgb, 255:red, 0; green, 0; blue, 0 }  ,fill opacity=1 ] (262.68,35.63) .. controls (262.68,34.87) and (263.29,34.25) .. (264.05,34.25) .. controls (264.81,34.25) and (265.43,34.87) .. (265.43,35.63) .. controls (265.43,36.38) and (264.81,37) .. (264.05,37) .. controls (263.29,37) and (262.68,36.38) .. (262.68,35.63) -- cycle ;
		\draw  [fill={rgb, 255:red, 0; green, 0; blue, 0 }  ,fill opacity=1 ] (110.63,162) .. controls (110.63,161.24) and (111.24,160.63) .. (112,160.63) .. controls (112.76,160.63) and (113.38,161.24) .. (113.38,162) .. controls (113.38,162.76) and (112.76,163.38) .. (112,163.38) .. controls (111.24,163.38) and (110.63,162.76) .. (110.63,162) -- cycle ;
		\draw  [fill={rgb, 255:red, 0; green, 0; blue, 0 }  ,fill opacity=1 ] (108.63,104) .. controls (108.63,103.24) and (109.24,102.63) .. (110,102.63) .. controls (110.76,102.63) and (111.38,103.24) .. (111.38,104) .. controls (111.38,104.76) and (110.76,105.38) .. (110,105.38) .. controls (109.24,105.38) and (108.63,104.76) .. (108.63,104) -- cycle ;
		\draw  [fill={rgb, 255:red, 0; green, 0; blue, 0 }  ,fill opacity=1 ] (222.96,162) .. controls (222.96,161.24) and (223.57,160.63) .. (224.33,160.63) .. controls (225.09,160.63) and (225.71,161.24) .. (225.71,162) .. controls (225.71,162.76) and (225.09,163.38) .. (224.33,163.38) .. controls (223.57,163.38) and (222.96,162.76) .. (222.96,162) -- cycle ;
		\draw    (112.75,210.5) -- (112.75,348) ;
		\draw [shift={(112.75,350)}, rotate = 270] [color={rgb, 255:red, 0; green, 0; blue, 0 }  ][line width=0.75]    (10.93,-3.29) .. controls (6.95,-1.4) and (3.31,-0.3) .. (0,0) .. controls (3.31,0.3) and (6.95,1.4) .. (10.93,3.29)   ;
		\draw  [color={rgb, 255:red, 155; green, 155; blue, 155 }  ,draw opacity=1 ][dash pattern={on 4.5pt off 4.5pt}] (45,63.5) -- (177,63.5) -- (177,215.5) -- (45,215.5) -- cycle ;
		\draw  [color={rgb, 255:red, 155; green, 155; blue, 155 }  ,draw opacity=1 ][dash pattern={on 4.5pt off 4.5pt}] (201,177) -- (365,177) -- (365,216) -- (201,216) -- cycle ;
		\draw  [color={rgb, 255:red, 155; green, 155; blue, 155 }  ,draw opacity=1 ][dash pattern={on 4.5pt off 4.5pt}] (267,244.5) -- (395,244.5) -- (395,329) -- (267,329) -- cycle ;
		
		\draw (60.17,74.5) node [anchor=north west][inner sep=0.75pt]   [align=left] {{\small Dense tanh (24)}};
		\draw (57.67,133) node [anchor=north west][inner sep=0.75pt]   [align=left] {{\small GRU ReLU (24)}};
		\draw (215.83,189.33) node [anchor=north west][inner sep=0.75pt]   [align=left] {{\small GRU ReLU (48)}};
		\draw (278.83,255.17) node [anchor=north west][inner sep=0.75pt]   [align=left] {{\small GRU ReLU (96)}};
		\draw (70,187.5) node [anchor=north west][inner sep=0.75pt]  [font=\footnotesize] [align=left] {{\fontsize{1.13em}{1.35em}\selectfont Dense sig. (1)}};
		\draw (285,301.5) node [anchor=north west][inner sep=0.75pt]  [font=\footnotesize] [align=left] {{\fontsize{1.13em}{1.35em}\selectfont Dense sig. (22)}};
		\draw (218.5,7.5) node [anchor=north west][inner sep=0.75pt]   [align=left] {{\scriptsize Input features (45)}};
		\draw (131.5,217) node [anchor=north west][inner sep=0.75pt]   [align=left] {{\fontsize{0.53em}{0.64em}\selectfont \textbf{Voice activity}}};
		\draw (367,180) node [anchor=north west][inner sep=0.75pt]   [align=left] {{\fontsize{0.53em}{0.64em}\selectfont \textbf{Noise spectral}}};
		\draw (374.25,190) node [anchor=north west][inner sep=0.75pt]   [align=left] {{\fontsize{0.53em}{0.64em}\selectfont \textbf{estimation}}};
		\draw (138.75,228) node [anchor=north west][inner sep=0.75pt]   [align=left] {{\fontsize{0.53em}{0.64em}\selectfont \textbf{detection}}};
		\draw (87.25,353.5) node [anchor=north west][inner sep=0.75pt]   [align=left] {{\fontsize{0.57em}{0.68em}\selectfont \textbf{VAD output (1)}}};
		\draw (293.75,354.5) node [anchor=north west][inner sep=0.75pt]   [align=left] {{\fontsize{0.57em}{0.68em}\selectfont \textbf{Gains outputs (22)}}};
		\draw (227.75,270.5) node [anchor=north west][inner sep=0.75pt]   [align=left] {{\fontsize{0.53em}{0.64em}\selectfont \textbf{Spectral }}};
		\draw (222.25,281) node [anchor=north west][inner sep=0.75pt]   [align=left] {{\fontsize{0.53em}{0.64em}\selectfont \textbf{subtraction}}};

	\end{tikzpicture}
	
	\caption{Deep Recurrent Neural Network Topology.}
	\label{figRNNtopology}
	
\end{figure}

As described in the original paper \cite{valin}, the network is so designed that it follows the usual structure of many conventional noise suppression algorithms. The basic idea behind the design of the system is that it can be divided into three subsystems: a Voice Activity Detector (VAD), a noise spectral estimation and a spectral subtraction block. Each subsystem includes a recurrent layer and specifically a gated recurrent unit (GRU).

Concerning VAD, it contributes significantly to the training process by helping the system differentiate noise from speech. It also outputs a voice activity probability even though it is not actively used in the inference process. 

To train our RNNoise and feature-extended systems we utilize the clean speech dataset included in the Edinburgh Dataset \cite{edinburgh} which is comprised of audio recordings, sampled at 48 kHz, of 28 English speakers (14 men and 14 women) with similar pronunciation. For noise recordings, we used a subset of the acoustical environments available in the DEMAND dataset \cite{demand}. These environments were then excluded from those used as the test set. The DEMAND dataset includes noise recordings corresponding to six distinct acoustic scenes (Domestic, Nature, Office, Public, Street and Transportation), which are further subdivided in multiple more specific noise sources \cite{demand}. Note that while we used clean speech and noise included in the Edinburgh Dataset, the samples used for training the systems are not the noisy samples found in the noisy speech subset of the Edinburgh Dataset, but rather samples mixed using the method described in \cite{valin}.

The test set used is the one provided in the Edinburgh  Dataset \cite{edinburgh}, which has been specifically created for speech enhancement applications and consists of wide-band (48kHz) clean and noisy speech audio tracks. The noisy speech in the set included four different SNR levels (2.5dB, 7.5dB, 12.5dB, 17.5dB). The clean speech tracks included in the set are recordings of two english language speakers, a male and a female. As for the noise recordings that were used in the mixing of the noisy speech tracks, those were derived from the DEMAND database \cite{demand} \cite{valentini}. More specifically, the noise profiles found in the testing set are:

\begin{itemize}
	\item \textbf{Living:} noise inside a living room (Domestic)
	\item \textbf{Office:} noise from a small office with three people using computers (Office)
	\item \textbf{Psquare:} noise from a public town square with many tourists (Street)
	\item \textbf{Cafe:} noise from the terrace of a cafe at a public square (Street)
	\item \textbf{Bus:} noise a public transit bus (Transportation)
\end{itemize}

The selection of the appropriate evaluation metrics is of great importance in the effort of regular evaluation of any system. In order to evaluate our system we used a metric that focuses on the sound quality (PESQ) and a metric that focuses on the intelligibility of the voice signal (STOI).

The wide-band Perceptual Evaluation of Speech Quality (PESQ) \cite{josh} is an objective and generally used standard for measuring sound quality. It takes account of features such as sound sharpness, speech volume, ambient noise, interruptions and interferences [35]. The PESQ scale calibration ranges from -0.5 to 4.5, with higher values corresponding to better quality.

The Short-Time Objective Intelligibility (STOI) \cite{taal} is a metric that increases according to the average intelligibility of the processed signal, given the original signal. Average intelligibility (or comprehensibility) is the percentage of words that are properly understood by a group of users. This metric ranges from 0 to 1.

\section{Results and Discussion}

It was deemed appropriate to present our results in a comparison between the reference RNNoise system and the modified version that makes use of the additional features. By comparing the two systems with regards to the PESQ quality metric, as seen in Fig.~\ref{figPESQ}, it becomes apparent that the modified version falls short by significant margin in all acoustic environment settings and in all SNR levels, but especially in higher SNRs. Similarly, examining the STOI intelligibility measure, as depicted in Fig.~\ref{figSTOI}, it is deduced that the modified version again falls severely short, but this time it is especially so for higher values of SNR. An exception to this appears to be the case of the "Living" audio scene, where, especially for low SNRs the performance of the two systems seems to be similar.

\begin{figure*}[htbp]
	\centering
	\begin{subfigure}[t]{0.48\linewidth}
		\centering
		\includegraphics[scale=0.24]{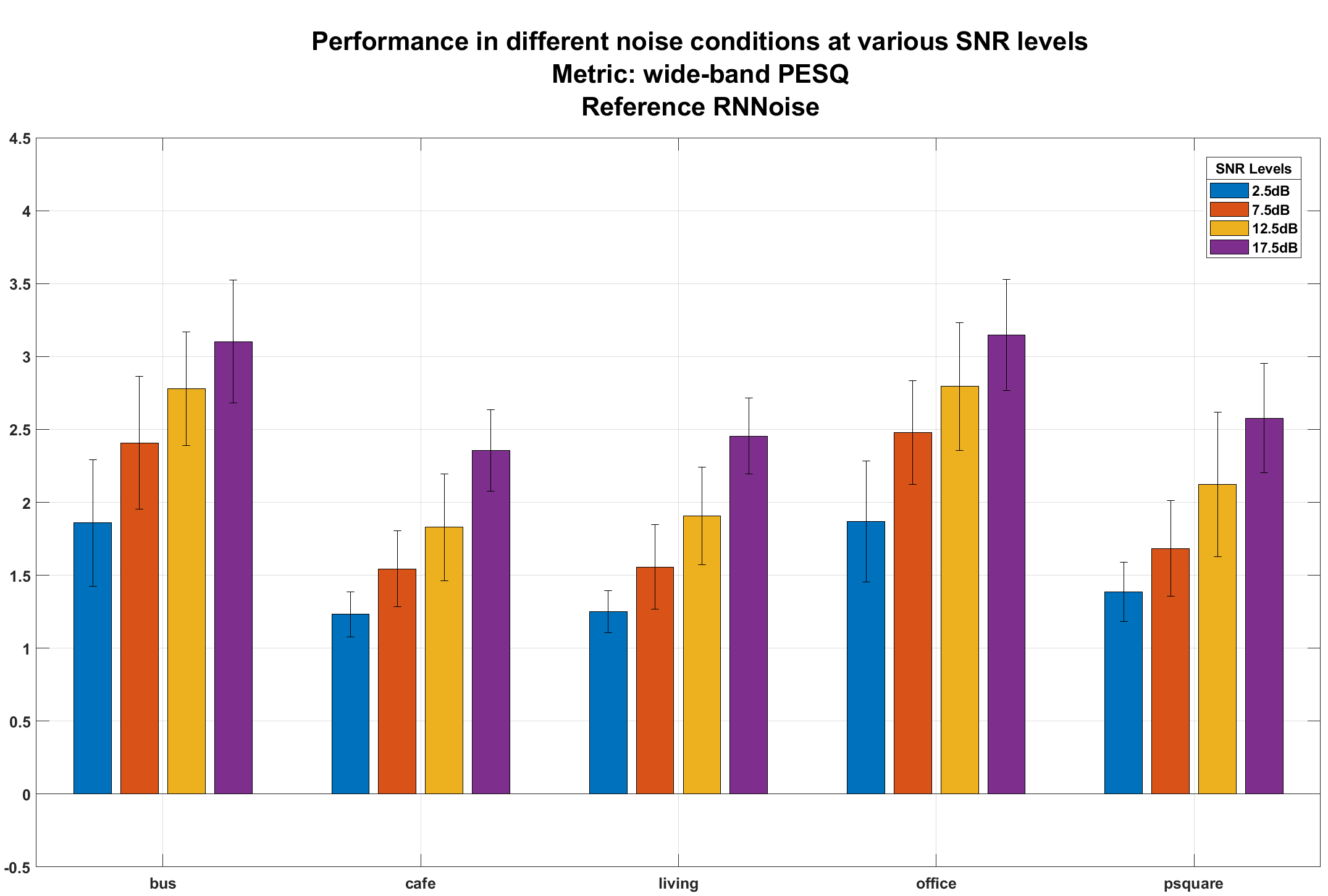}
		\caption{}
	\end{subfigure}
~
	\begin{subfigure}[t]{0.48\linewidth}
		\centering
		\includegraphics[scale=0.24]{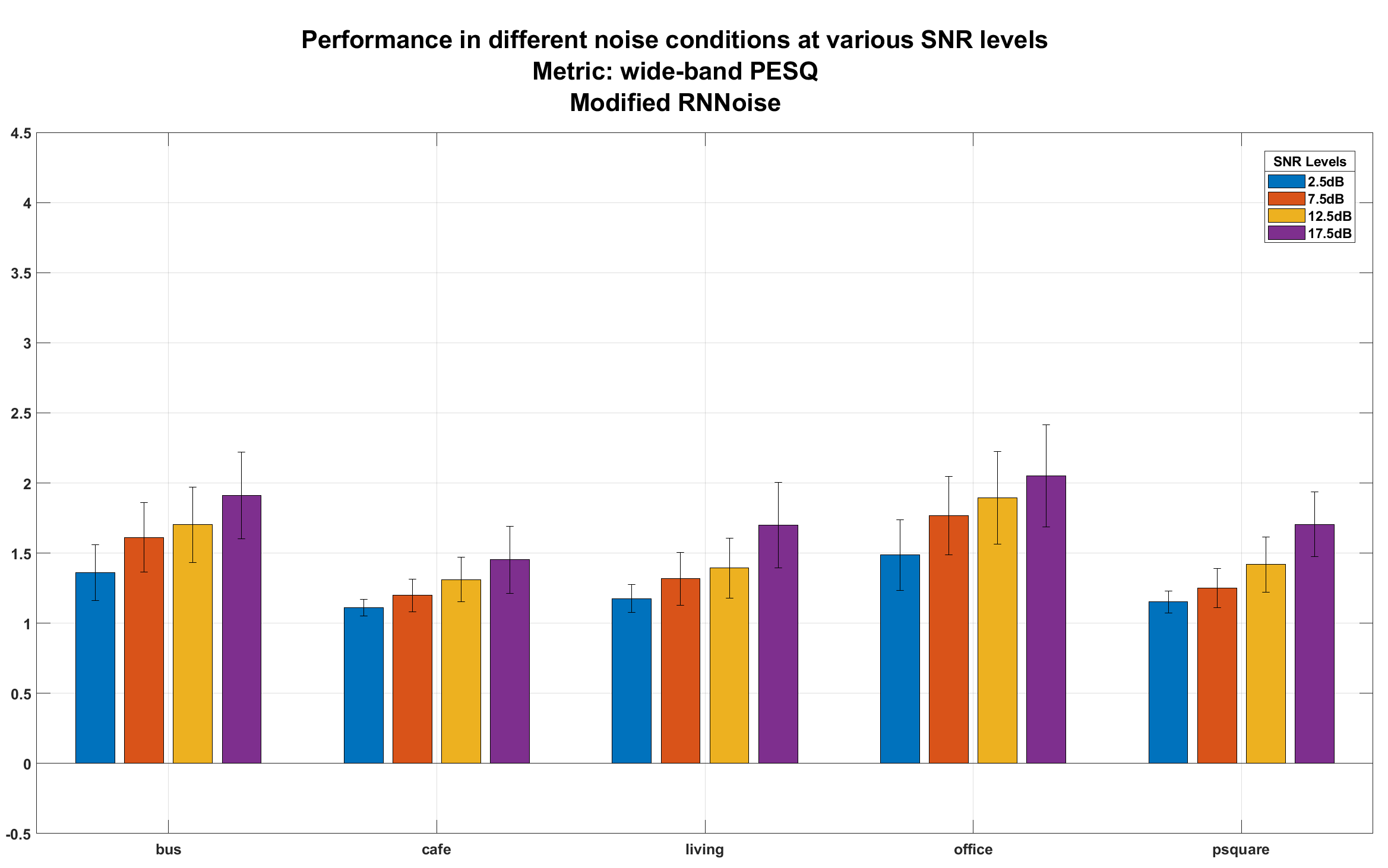}
		\caption{}
	\end{subfigure}

	\caption{Extended and Reference System PESQ performance in different acoustical environments under various SNR levels: \\ a. Reference system and b. Extended system}
	\label{figPESQ}
\end{figure*}

\begin{figure*}[htbp]
	
	\begin{subfigure}{0.48\textwidth}
		\centering
		\includegraphics[scale=0.24]{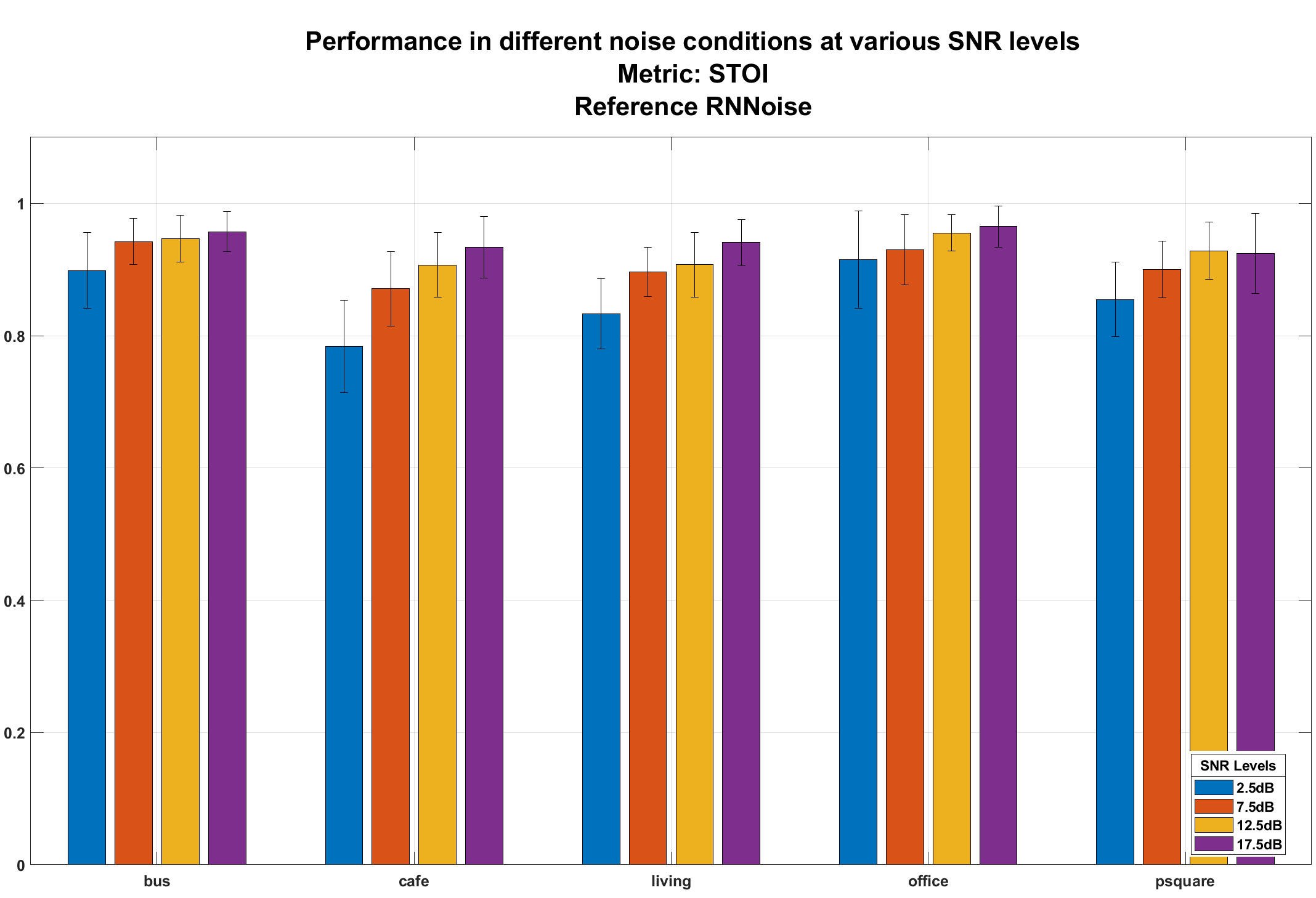}
		\caption{}
	\end{subfigure}
	~
	\begin{subfigure}{0.48\textwidth}
		\centering
		\includegraphics[scale=0.24]{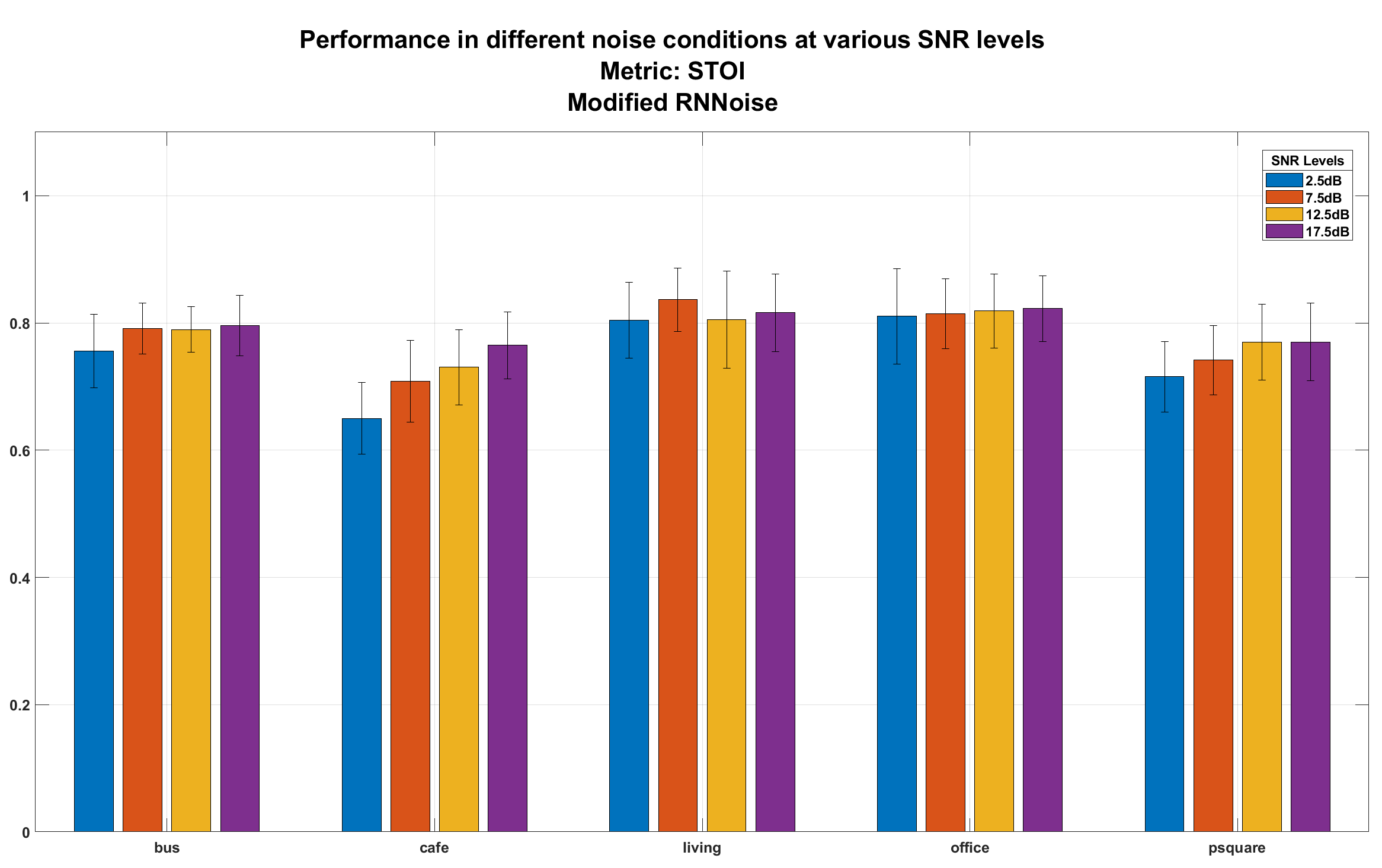}
		\caption{}
	\end{subfigure}
	\caption{ Extended and Reference System STOI performance in different acoustical environments under various SNR levels: \\ a. Reference system and b. Extended system}
	\label{figSTOI}
\end{figure*}

Overall, the modified system appears to have a generally worse performance than the reference version of RNNoise. Having also compared several pairs of spectrograms of both denoising system cases, it was observed that in general the modified version does indeed subtract less noise components.

Having taken these results into consideration, we now discuss some avenues for further future development that will hopefully yield better performance results. 

Firstly, while for the base system \cite{valin} Valin notes that adding more hidden layers does not improve performance significantly, we believe that it might indeed be beneficial for our extended system. Given that we provide the system with more and diverse input information, the RNN might be able to better exploit the proposed features with additional hidden layers.

Currently our extended system utilizes the additional features as calculated on the full spectrum of each window processed by the RNN. We believe that the system's performance can potentially be improved by calculating these features for each individual subband of the modified Bark scale or for a small selection of them. This will subsequently lead to an increase of the RNN's input features and as such the network's hidden layers will have to be adapted to properly accommodate this change.

Studying samples processed by our extended system, we speculate that the system could benefit from changing how aggressively the noise suppression occurs. This can be achieved by fine-tuning the value of the $\gamma$ parameter in the loss function \eqref{eqLoss}, keeping in mind that smaller $\gamma$ values lead to more aggressive suppression. According to \cite{valin}, setting $\gamma = \frac{1}{2}$ is an optimal balance.

We initially considered also using Root Mean Square (RMS), which is related to the signal's energy and its change over time, and Spectral Flatness which is used to discern tone-like from noise-like signals. However, when we calculated and visualized these features for our dataset, we discovered that they offered limited variance and had many outliers. This led us to omit them from our feature set as we believed that they would increase input dimensionality more than would benefit performance. Revisiting these features under the subbanding context described above might prove to improve the system.

Finally, we believe that further research can be done regarding the performance of the base and extended systems as the training dataset increases in size and diversity.

\section{Conclusion}
In this paper we have presented our efforts to extend and improve a hybrid speech enhancement system. We proposed features which we believed would further assist the denoising process and assessed them as inputs to the recurrent neural network. We illustrated our toolchain for training the system with extended input features and compared the system against a reference RNNoise instance trained using the same training parameters. We discussed our findings from this process, concluding that the extra features have no obvious positive effect on the system's performance for the training test size used. Finally, we laid out our thoughts on future avenues to be explored for further improvement of the base system using spectral features.

\section*{Acknowledgment}

We would like to thank our teachers, Dr. Charalampos A. Dimoulas (Associate Professor) and Dipl. Iordanis Thoidis (PhD candidate) (Laboratory of Electroacoustics and TV Systems, School of Electrical and Computer Engineering, Aristotle Univeristy of Thessaloniki) for their enthusiastic guidance and support throughout the research process and writing of this paper.


\begin{thebibliography}{00}
	
\bibitem{valin} J.-M. Valin, ``A Hybrid DSP/Deep Learning Approach to Real-Time Full-Band Speech Enhancement,'' International Workshop on Multimedia Signal Processing, 2018.

\bibitem{lai} Y.-H. Lai et al. “Deep Learning-Based Noise Reduction Approach to Improve Speech Intelligibility for Cochlear Implant Recipients,” Ear and hearing vol. 39,4 (2018): 795-809.

\bibitem{donahue}C. Donahue, B. Li and R. Prabhavalkar, ``Exploring Speech Enhancement with Generative Adversarial Networks for Robust Speech Recognition,'' 2018 IEEE International Conference on Acoustics, Speech and Signal Processing (ICASSP), Calgary, AB, Canada, 2018, pp. 5024-5028.
	
\bibitem{weiner} J. S. Lim and A. V. Oppenheim, ``Enhancement and bandwidth compression of noisy speech,'' in Proceedings of the IEEE, vol. 67, no. 12, pp. 1586-1604, Dec. 1979.

\bibitem{welchKalman} G. Welch et al. ``An Introduction to the Kalman Filter,'' Proc. Siggraph Course 8, 2006.
\url{https://www.researchgate.net/publication/200045331_An_Introduction_to_the_Kalman_Filter}

\bibitem{wang}D. Wang, J. Chen, ``Supervised Speech Separation Based on Deep Learning: An Overview,'' in IEEE/ACM Transactions on Audio, Speech, and Language Processing, vol. 26, no. 10, pp. 1702-1726, Oct. 2018

\bibitem{shifas} M. Shifas, N. Adiga, V. Tsiaras, Y. Stylianou, ``A non-causal FFTNet architecture for speech enhancement,'' Interspeech, 2019.

\bibitem{hybrid1} Y.-H. Tu, I. Tashev, S. Zarar, C.-H. Lee, ``A Hybrid Approach to Combining Conventional and Deep Learning Techniques for Single-Channel Speech Enhancement and Recognition,'' 2531-2535, 10.1109/ICASSP.2018.8461944, Apr. 2018.

\bibitem{hybrid2} M. J. Coto, J. C. Goddard, L. Di Persia, H. L. Rufiner, ``Hybrid Speech Enhancement with Wiener Filters and Deep LSTM Denoising Autoencoders,''
1-8, 10.1109/IWOBI.2018.8464132.

\bibitem{opus} J.-M. Valin, G. Maxwell, T. B. Terriberry, and K. Vos, ``High-quality,
low-delay music coding in the Opus codec,'' in Proc. 135th AES Convention, 2013.

\bibitem{giannakopoulos} T. Giannakopoulos and A. Pikrakis, "Introduction to Audio Analysis: a MATLAB Approach," Academic Press Is an Imprint of Elsevier, 2014. 

\bibitem{andersson} T. Andersson, "Audio Classification and Content Description," MS Thesis, Luleå University of Technology, 2004.

\bibitem{maningo} E. Maningo, “Understanding What Does RMS Stands for in Audio: Definition \& Details,” Audio Recording, 2012.

\bibitem{peeters} G. Peeters, “A large set of audio features for sound description (similarity and classification) in the Cuidado Project,” Institut de Recherche	et Coordination Acoustique/Musique (IRCAM), 2004.

\bibitem{dubnov} S. Dubnov, “Generalization of Spectral Flatness Measure for Non-Gaussian Linear Processes,” IEEE Signal Processing Letters, vol. 11, no. 8, 2004.

\bibitem{sox} C. Bagwell, ``SoX - Sound eXchange,'' \url{http://sox.sourceforge.net/Main/HomePage}, 2015.

\bibitem{tensorflow} M. Abadi, A. Agarwal, P. Barham, E. Brevdo, Z. Chen, C. Citro et al., ``Tensorflow: A system for large-scale machine learning,'' \url{https://www.tensorflow.org}, 2015.

\bibitem{josh}
J. O’Farrell,  ``What Is: PESQ?'' [Blog] Transforming global communications, 2020. \url{https://www.spearline.com/blog/post/what-is--pesq-/} (Accessed: 04 June 2020)

\bibitem{taal}
C. H. Taal et al. ``A short-time objective intelligibility measure for time-frequency weighted noisy speech,'' 2010 IEEE International Conference on Acoustics, Speech and Signal Processing, Dallas, TX, USA, 14-19 March 2010. IEEE, 2010, pp. 4214-4217. IEEE Xplore,
\url{https://ieeexplore.ieee.org/abstract/document/5495701}

\bibitem{edinburgh}C. Valentini-Botinhao, ``Noisy speech database for training speech enhancement algorithms and TTS model,'' University of Edinburgh, School of Informatics, Centre for Speech Technology Research (CSTR), 2017, 2016 [sound].
\url{https://doi.org/10.7488/ds/2117}

\bibitem{demand}
J. Thiemann, N. Ito and E. Vincent, ``DEMAND: a collection of multi-channel recordings of acoustic noise in diverse environments,'' (Version 1.0) [Data set], Presented at the 21st International Congress on Acoustics (ICA 2013), Montreal, Canada: Zenodo, 2013.
\url{http://doi.org/10.5281/zenodo.1227121}

\bibitem{valentini}
C. Valentini-Botinhao, X. Wang, S. Takaki, J. Yamagishi, ``Speech Enhancement for a Noise-Robust Text-to-Speech Synthesis System Using Deep Recurrent Neural Networks,'' INTERSPEECH, 2016.

\end{thebibliography}
\end{document}